# Spatiotemporal Assessment of $So_2$, $SO_4$ and AOD from over MENA Domain from 2006 – 2016 Using multiple satellite Data and Reanalysis MERRA-2 Data

Muhammed ELTAHAN, Mohammed Magooda ,

Aerospace Engineering Department, Cairo University, Cairo, Egypt

Corresponding author: Muhammad Eltahan (Muhammedsamireltahan@gmail.com)

## Abstract

The sulfur pollutants are the source of a sizeable portion of the air pollution. In this work, the spatial distribution and temporal trend of the mass concentration of two of the critical sulfur pollutants, $SO_2$ and $SO_4$, in addition to the aerosol optical properties (AOD) were analyzed over the region of the Middle East and North Africa (MENA) from multiple satellite recourses and Modern Era-Retrospective Analysis for Research and Applications version 2 (MERRA-2) reanalysis data.

The $So_2$ and $So_4$ data used in these analyses are obtained from (MERRA-2) with a resolution of 0.5°×0.625° throughout a period of 10 years (2005 – 2015). On the other hand, the temporal trend and spatial distribution of AOD was identified from four different satellite data. (1) moderate-resolution imaging spectroradiometer (MODIS) Level 3 AOD data at 550 nm wavelengths from Collection 6 algorithm (combined dark target and deep blue algorithms) are used for 10 years temporal analysis (2006-2015). Data were obtained for the period 2006-2015 (2) Multi-angle imaging spectroradiometer (MISR) with 0.5 deg spatial resolution for the same 10 years (2006-2015). (3) Sea-Viewing Wide Field-of-View Sensor (SeaWIFS) with 0.5 deg for the period (2005-2010). (4) Ozone Monitoring Instrument (OMI) AOD at 500 nm wavelength with resolution 1 degree

**Keywords:** MODIS, MISR, OMI, SeaWIFS , MERRA-2

# Introduction

Air pollutants affect almost every aspect of life, many of them have direct impact on public health and agriculture, aside from the greenhouse effects that has further consequences. $SO_2$ is considered a primary irritant and is known for having sever short and long-term health effects. It affects not only the respiratory system, but the eyes and skin as well, while lab experiments show that it can affect other systems. (Maynard D. etal 2007, Nan S. 2010, CAS 7446-09-5, 2004). The effect of such pollutants is dose dependent, and long-term exposure might cause reactive air ways dysfunction syndrome whiles some studies associates puts Sulfur dioxide as a possible carcinogenesis when acting with other substances and as a contributor in developing Ischemic stroke in the brain. While Sulfates $SO_4$ can form a sizable part of fine particulate matter, which contributes greatly in poisoning rain and consequently surface water and soil, besides the fact that these acid rains damage forests and the ecosystem in general (Maynard D. etal 2007, Nan S. 2010, CAS 7446-09-5, 2004)

Sulfur dioxide ($SO_2$) gas is mainly generated as a byproduct of fuel's sulfur oxidation during burning processes. It is emitted from stationary sources such as power plants, different industries or mobile sources such as vehicles. Noting that the maximum annual allowable limit in the Executive Regulation of Law no. 4/1994 amended by law no. 9/2009 is 60 μg/m3(State of Environment Report- 2010). However, Sources of $SO_4$ are the ocean, the soil, and the oxidation of the gaseous compounds.( C. E. Junge , 1960). While the aerosol optical depth (AOD) is one of the main physical indicators Quantifying the atmospheric turbidity and air pollution (Wenmin Qin ,.etal 2018). It was found that the Total AOD roughly can be estimated approximately to be summation of DUAOD, BCAOD, OCAOD, SSAOD and SO4AOD values in addition to constant. It was clear that SO4AOD is the main driving factor for the spatial distribution of total AOD values over mainland China (Wenmin Qin ,.etal 2018).

It was examined annual changes in OMI NO2 and SO2 columns over the Middle East over 2005–2014, reporting that, after increases in the period 2005–2010, there was a reduction in these gases, either due to new regulatory legislation or due to falls in economic output associated with regional conflicts and geopolitical controls Lelieveld et al. (2015). Detailed times series analysis for $So_2$ over 1000 locations (focusing on urban, oil refinery, oil port, and power plant targets) over the Middle East from 2005 to 2014 using OMI data was established. Very few SO2 trends were detected, which varied in direction and magnitude (23 increasing and 9 decreasing). (Michael P. Barkley ..etal 2017). similar regional-scale study by Krotkov et al (2016) also observed that during 2005–2008 OMI SO2 columns dropped by 20 % after 2010, only recovering to 2005 levels in 2014.

The Spatial and temporal trends of AOD and their main constitutions were studied. The contributions of dust and So$_4$ to AOD Total were investigated using MODIS and MERRA-2 aerosol reanalysis dataset for 38 years (1980-2017) over china. The result showed that dust aerosol optical depth ( DUAOD ) and So4 aerosol optical depth (SO4AOD) were the main driving factors for the distributions of the total AOD values in different climate zones over mainland China with high contributions. The annual mean AOD for, DUAOD and SO4AOD were 25.43%, 49.51%, respectively (Wenmin Qin…etal,2018)

The region of the Middle East and the Mediterranean is recognized to be one of the most effective areas in the world's climate changes (Giorgi and Lionello, 2008; Hoerling et al., 2012; Lelieveld et al., 2012; Hemming et al., 2010); The number of days with a remarkably high temperature in that area increased greatly, at the same time the number of cool days has dropped expressively (IPCC, 2014). The studies predict that temperature will increase more while precipitation will decrease in that region. In addition to the middle east's climatic importance, it also represents an outstanding importance from an atmospheric chemistry point of view [Klingmüller.K,etal 2016] as it falls at the center of what is known as the dust-belt (Astitha et al., 2012). Multiple sources of atmospheric soil dust were recognized by (Prospero, et al., 2002) in the MENA domain. The high concentrations of atmospheric dust near the Earth's surface is the main factor in the high levels of the aerosol optical depth (AOD) (e.g., Hsu et al., 2012). Studies show that mineral dust, which is a major component of atmospheric aerosols, has great effects on several aspects on the environmental system (Knippertz et al., 2014). Other studies have shown that the region of the Middle east has witnessed great increases in the AOD levels throughout the last ten years. These studies were conducted using several numerical models (Pozzer et al., 2015) along with observational data using remote sensing, e.g., from SeaWiFS ( IPCC, 5 2014 ; Hsu et al., 2012) MISR, MODIS and AERONET (de Meij et al., 2012; Stevens and Schwartz, 2012; de Meij and Lelieveld, 2011).

Many studies were conducted in order to evaluate AOD retrieval from different satellite sensors. On global scale CALIPSO AOD was underestimated with respect to MODIS AOD (X.Ma, etal.2013). Globally, for 13 year (1997-2010 ), approximately three fourths of the comparisons performed using both AERONET and SeaWiFS and AOD at 550 nm resulted in an absolute difference of 0.05 + 0.2 τA,550, when considering only SeaWiFS data with quality flag (QA = 3), which falls within the expected margin of error, (Sayer ,A. M., etal, 2012).

The AOD ratio of MODIS to MISR is greater than 1 over East and Central Asia, Indonesia, Western states of the US and South America,. While the ratio is significantly less than 1 over the East coast of South America, East Coast of South Africa, west of Australia and the Arabian Peninsula (Shi, Y,2011).  The results obtained by pairing the adjacent MISR and MODIS pixels confirmed the findings in earlier studies, performed over the mainland in south east Asia, which stated that both are highly correlated. Other prior studies stated that red and near infrared bands of MISR are brighter with respect to independent standards by 3% and 1%, respectively ( Abdou et al., 2005; Kahn et al., 2005b; Bruegge et al., 2003) ( over Karachi, Lahore, Jaipur, and Kanpur between 2007 and 2013 ).

Generally, data obtained from MODIS STD and AERONET observations highly agree when compared over bright surfaces such as desert or coastal sites (e.g., over Jaipur or Karachi), while comparing MISR retrievals with AERONET observations over regions close to the ocean (e.g., over Karachi) showed a high degree of correlation. Comparing data from MODIS Deep Blue (DB) and AERONET observations showed a reasonable agreement over all sites, as did OMI retrievals. (Bibi,H, etal 2015). During MACC-II 2007–2008 reanalysis over Northern Africa and Middle East, the agreement between MISR, OMI, MODIS and MACC-II is, in overall, rather good, repeating the same AOD spatial distributions (E. Cuevas et al.,2015)

In the current work, we present analysis of the spatial and temporal trands of So2 and So4 in addition comparative study of spatial and temporal AOD over MENA) domain using four satellite data MODIS, MISR,Sea-viewing Wide Field-of-view Sensor (SeaWiFs) and Ozone Monitoring Instrument (OMI) over the MENA domain. Section 2 presents the different data sources that had been used in the current work, section 3. discusses the results and section 4 will present the conclusions and future work.

## 2 Data

Since meteorological, chemical, biological...et. Observations became available, they play critical role in understanding Earth systems, in addition to their role for validating Numerical Models output. They can be assimilated to the numerical models to improve the output estimation.

This section provides highlight on observational data that used in current study from different sensors MODIS, MISR, OMI and SEAWIFS to investigate the spatial distribution and Temporal trend of AOD. In addition to the both sulfate pollutants from MERRA-2.

### 2.1 MODIS

The scientific instrument moderate-resolution imaging spectroradiometer (MODIS) was on board of the Terra and Aqua satellites in 1999 and 2002 respectively. Many and different products were retrieved from MODIS. Its aerosol products monitor optical properties like aerosol optical depth (AOD) and single scattering albedo (SSA) over

Land and Oceans.Level-3 data with resolution (1º degree) from MODIS is used in the benchmark over the MENA domain that conducted to evaluate AOD product. Collection-6 of aerosol product is used for AOD Calculations, that merges together both algorithms of dark target (DT) and deep blue (DB).

## 2.2 MISR

MISR instrument on the Terra satellite is used to capture several angular views of the data of the atmosphere and surface (Diner et al., 1998). These views are used to salvage multiple physical aerosol parameters (Moroney et al 2002), they are also used to capture some aerosol microphysical properties (Martonchik et al., 2004). More detailed information about the aerosol algorithm and how to retrieve them can be found in Diner et al. (2001)

For aerosol optical depth, we used monthly average Global 0.5 × 0.5 Degree Aerosol Product (MIL3MAE) with Aerosol optical Depth AOD 555 nm to investigate the temporal analysis in addition to the spatial distribution.

## 2.3 SeaWiFs

In order to capture global ocean biological data, Satellite-borne Sensor, SeaWIFS (SeaViewing Wide Field-of-View Sensor) was built .This sensor has been around for 13 year (1997 to 2010), Its main mission was to identify the amount of produced chlorophyll by marine phytoplankton. This sensor recorded the information in 8 optical bands

The SeaWiFS (SWDB_L3M05) Level-3 product contains monthly global gridded (0.5 x 0.5 deg) Version 4 data derived from SeaWiFS Deep Blue Level 3 daily gridded data at the same spatial resolution. The primary data that is used aerosol optical thickness at 550nm for long term temporal and spatial distribution

## 2.4 OMI

In July 2004, OMI was launched on NASA's EOS-Aura satellite, also part of the A-train 15 constellation. In this study we analyze the OMI (OMAERUVd-v003) product to plot long term time series of Aerosol Optical Depth at 500 nm.In 2008 NASA Goddard Earth Sciences Data and Information Services Center (GES DISC) released OMAERUVd level-3 daily global gridded data with (1x1 deg) horizontal resolution, which is based on level 2 Aerosol data product OMAERUV that is originally an improved version of TOMS version-8 algorithm. http://disc.gsfc.nasa.gov/Aura/OMI/omaeruvd_v003.shtml

## 2.5 MERRA-2

In this work the data provided through MERRA-2 -Modern-Era Retrospective Analysis for Research and Applications version 2- were used MEERA-2 is an update to its proceeding intended to provide an ongoing climate analysis that goes beyond the MERRA's terminus, while addressing it's known limitations

Unlike in MERRA, all data collections from MERRA-2 are provided on the same horizontal grid. This grid has 576 points in the longitudinal direction and 361 points in the latitudinal direction, corresponding to a resolution of 0.625°×0.5°. The longitudinal resolution of the data is changed from 0.667° in MERRA and the latitudinal resolution remains unchanged (0.5°).

One of the recent studies that use MERRA-2 and shows that regionally averaged time series of the ModerateResolution Imaging Spectroradiometer (MODIS) observed CDNC of low, liquid-topped clouds is well predicted by the MERRA2 reanalysis near-surface sulfate mass concentration over decadal timescales. A multiple linear regression between MERRA2 reanalyses masses of sulfate (SO4), black carbon (BC), organic carbon (OC), sea salt (SS), and dust (DU) shows that CDNC across many different regimes can be reproduced by a simple power-law fit to near-surface SO4, with smaller contributions from BC, OC, SS, and DU.( Daniel T. McCoy, 2018).

## 3- Results and Discussions

The results in this section are divided under two main subsections, the first subsection handle the spatial and temporal trend of AOD form different satellite sensors over the MENA domain. The second subsection highlight the spatial and temporal trend of Sulfate pollutant over MENA domain from MERRA-2 data set.

### 3.1 Aerosol Optical Depth (AOD)

Comparison between spatial distributions of seasonal average of AOD from MODIS and MISR for the 10 years period is shown Figure1. Both sensors provide almost the same distribution pattern. Summer and spring seasons (JJA, MAM) have the highest spatial AOD while autumn season (SON) has the lowest AOD. The maps reveal five areas of high AOD that agree with the areas of global dust sources defined by (Propero, etal 2002). The areas are (1) Lake Chad Basin and the Bodele Depression, (2) Saudi Arabia peninsula (3)Tunisia & Northeast Algeria,(4) Ethiopia and Eritrea. The areas are marked in the figure according to the same sequence mentioned in the previous line.

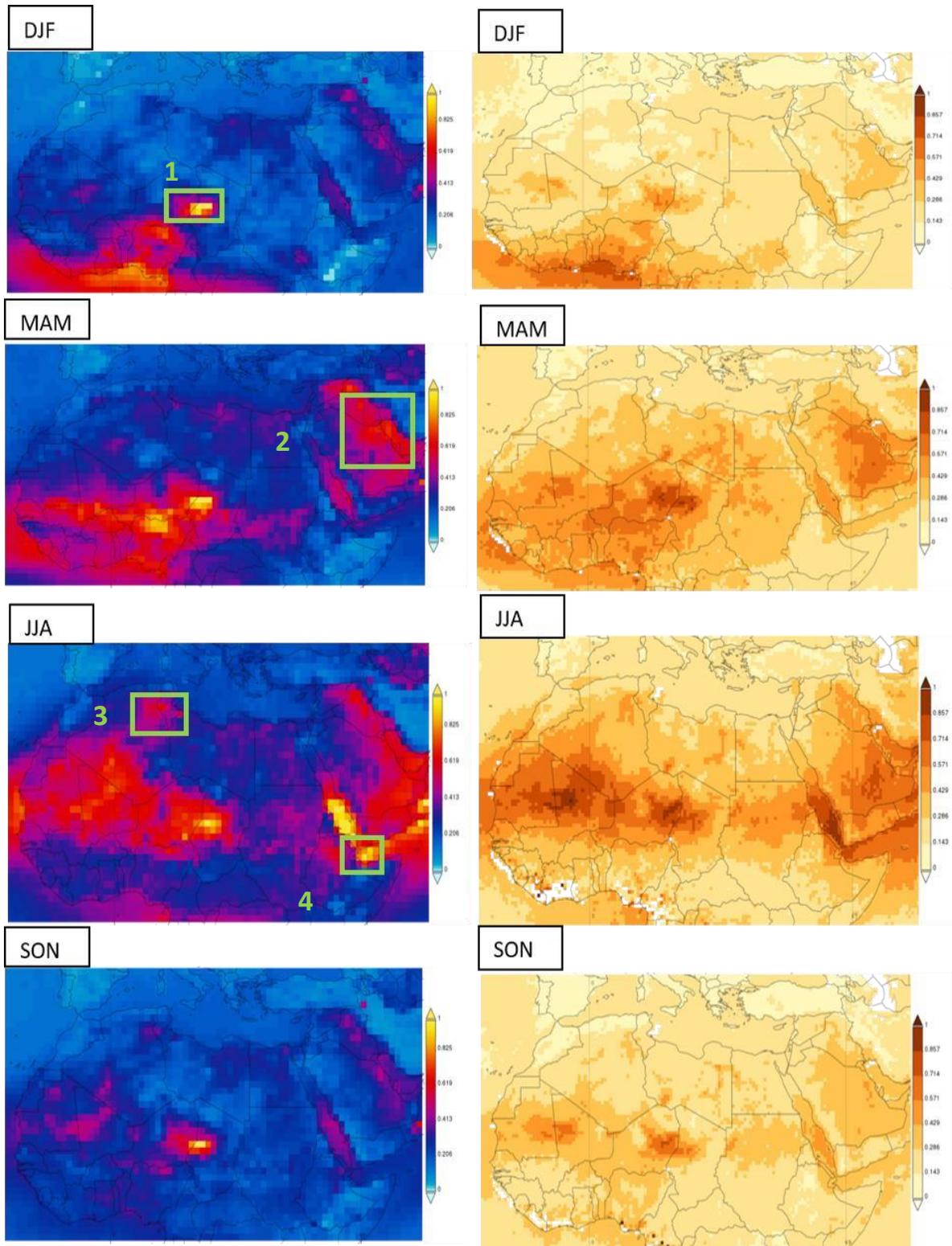

Figure 1 Seasonal average of AOD (2006 -2015) from both MODIS [Right panel] and MISR [Left Panel]

The inter-annual seasonal time series from both MODIS and MISR are shown in Figure 2. Quantitatively speaking, the highest AOD is found in the summer (JJA) 2011 which observed by MISR (0.4149) and MODIS (0.422) while the lowest average AOD was (0.2125) observed by MODIS at autumn (SON) 2013, although that MISR observed it higher (0.2298). In general, it is shown that both summer and spring have higher average of AOD than both winter and autumn from the both sensors

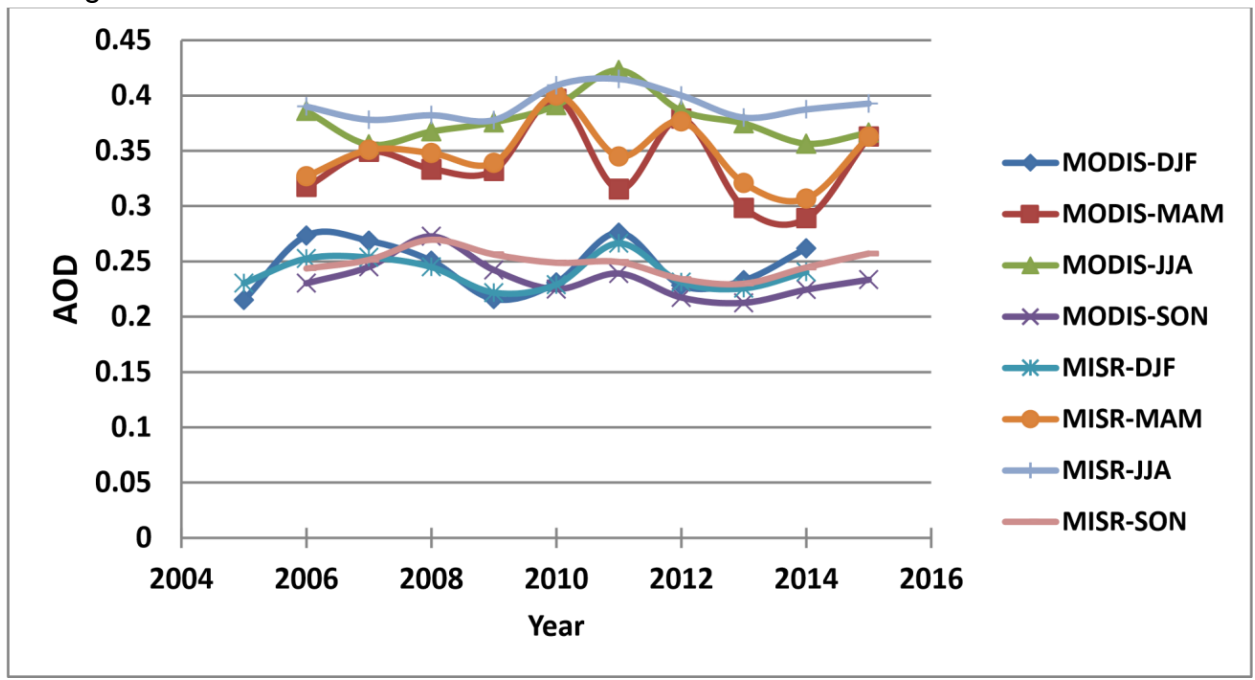

Figure 2 Inter-annual time series of AOD Monthly (2006 -2015) from both MODIS and MISR

Inter-annually time series monthly average AOD ten years (2006 -2015) from both sensors MODIS and MISR are shown in Table1.For MODIS, the highest monthly average AOD was in June and July 2011(0.45 and 0.459) and the next highest peaks were in March 2012 and 2010 (0.441 and .442). The lowest monthly average was in November 2013 (0.168). The max and mini average AOD for every month with corresponding year are shown in Table 2

Inter-annually time series monthly average AOD ten years (2006 -2015) from both sensors MODIS and MISR are shown in Table 1. For MODIS, The highest monthly average AOD was in June and July 2011(0.45 and 0.459) and the next highest peaks were in March 2012 and 2010 (0.441 and .442). The lowest monthly average was in November 2013 (0.168). The max and mini average AOD for every month with corresponding year are shown in Table 2. The absolute difference between AOD from MODIS and MISR is presented in Table 3

Table 1: Ten years (2005-2016) Monthly average of time series of AOD over MENA domain from MISR [Upper Table] and MODIS [Lower Table]

| Year/Month | Jan. | Feb | Mar. | April | May | June | July | Aug. | Sep. | Oct. | Nov. | Dec. |
|---|---|---|---|---|---|---|---|---|---|---|---|---|
| 2006 | 0.204 | 0.261 | 0.342 | 0.319 | 0.289 | 0.416 | 0.381 | 0.3615 | 0.274 | 0.222 | 0.194 | 0.2437 |
| 2007 | 0.301 | 0.275 | 0.366 | 0.353 | 0.328 | 0.361 | 0.366 | 0.3399 | 0.285 | 0.242 | 0.208 | 0.2042 |
| 2008 | 0.242 | 0.36 | 0.301 | 0.398 | 0.302 | 0.37 | 0.39 | 0.3418 | 0.328 | 0.286 | 0.204 | 0.2088 |
| 2009 | 0.233 | 0.31 | 0.356 | 0.31 | 0.33 | 0.368 | 0.418 | 0.342 | 0.29 | 0.226 | 0.21 | 0.1816 |
| 2010 | 0.21 | 0.257 | 0.442 | 0.382 | 0.366 | 0.401 | 0.413 | 0.3592 | 0.258 | 0.235 | 0.182 | 0.2018 |
| 2011 | 0.25 | 0.241 | 0.283 | 0.336 | 0.327 | 0.45 | 0.46 | 0.3583 | 0.257 | 0.259 | 0.201 | 0.2316 |
| 2012 | 0.252 | 0.345 | 0.442 | 0.346 | 0.35 | 0.417 | 0.388 | 0.3533 | 0.25 | 0.218 | 0.184 | 0.1906 |
| 2013 | 0.229 | 0.265 | 0.272 | 0.311 | 0.313 | 0.41 | 0.382 | 0.3315 | 0.262 | 0.209 | 0.167 | 0.2022 |
| 2014 | 0.216 | 0.281 | 0.295 | 0.291 | 0.281 | 0.361 | 0.373 | 0.3356 | 0.275 | 0.213 | 0.185 | 0.2168 |
| 2015 | 0.296 | 0.273 | 0.346 | 0.427 | 0.315 | 0.402 | 0.347 | 0.3498 | 0.282 | 0.23 | 0.189 | 0.2786 |

| Year/Month | Jan. | Feb | Mar. | April | May | June | July | Aug. | Sep. | Oct. | Nov. | Dec. |
|---|---|---|---|---|---|---|---|---|---|---|---|---|
| 2006 | 0.204 | 0.261 | 0.342 | 0.319 | 0.289 | 0.416 | 0.381 | 0.3615 | 0.274 | 0.222 | 0.194 | 0.2437 |
| 2007 | 0.301 | 0.275 | 0.366 | 0.353 | 0.328 | 0.361 | 0.366 | 0.3399 | 0.285 | 0.242 | 0.208 | 0.2042 |
| 2008 | 0.242 | 0.36 | 0.301 | 0.398 | 0.302 | 0.37 | 0.39 | 0.3418 | 0.328 | 0.286 | 0.204 | 0.2088 |
| 2009 | 0.233 | 0.31 | 0.356 | 0.31 | 0.33 | 0.368 | 0.418 | 0.342 | 0.29 | 0.226 | 0.21 | 0.1816 |
| 2010 | 0.21 | 0.257 | 0.442 | 0.382 | 0.366 | 0.401 | 0.413 | 0.3592 | 0.258 | 0.235 | 0.182 | 0.2018 |
| 2011 | 0.25 | 0.241 | 0.283 | 0.336 | 0.327 | 0.45 | 0.46 | 0.3583 | 0.257 | 0.259 | 0.201 | 0.2316 |
| 2012 | 0.252 | 0.345 | 0.442 | 0.346 | 0.35 | 0.417 | 0.388 | 0.3533 | 0.25 | 0.218 | 0.184 | 0.1906 |
| 2013 | 0.229 | 0.265 | 0.272 | 0.311 | 0.313 | 0.41 | 0.382 | 0.3315 | 0.262 | 0.209 | 0.167 | 0.2022 |
| **2014** | 0.216 | 0.281 | 0.295 | 0.291 | 0.281 | 0.361 | 0.373 | 0.3356 | 0.275 | 0.213 | 0.185 | 0.2168 |
| **2015** | 0.296 | 0.273 | 0.346 | 0.427 | 0.315 | 0.402 | 0.347 | 0.3498 | 0.282 | 0.23 | 0.189 | 0.2786 |

Table 2: Max and Mini. Monthly average of AOD for Ten years (2005-2016) over MENA domain from MISR [Upper Table] and MODIS [Lower Table]

| | Jan. | Feb | Mar. | April | May | June | July | Aug. | Sep. | Oct. | Nov. | Dec. |
|---|---|---|---|---|---|---|---|---|---|---|---|---|
| **Year** | 2007 | 2012 | 2010 | 2010 | 2010 | 2010 | 2011 | 2011 | 2008 | 2011 | 2009 | 2015 |
| **Max** | 0.267 | 0.327 | 0.399 | 0.4106 | 0.39 | 0.42133 | 0.44742 | 0.392 | 0.3519 | 0.2636 | 0.2148 | 0.2383 |
| **Year** | 2014 | 2011 | 2013 | 2014 | 2006 | 2009 | 2015 | 2013 | 2012 | 2013 | 2013 | 2009 |
| **Min** | 0.216 | 0.244 | 0.285 | 0.3064 | 0.3165 | 0.35965 | 0.3722 | 0.352 | 0.2861 | 0.2153 | 0.178 | 0.1929 |

| | Jan. | Feb | Mar. | April | May | June | July | Aug. | Sep. | Oct. | Nov. | Dec. |
|---|---|---|---|---|---|---|---|---|---|---|---|---|
| **Year** | 2007 | 2008 | 2010 | 2015 | 2010 | 2011 | 2011 | 2006 | 2008 | 2008 | 2009 | 2015 |
| **Max** | 0.301 | 0.36 | 0.442 | 0.427 | 0.366 | 0.45 | 0.46 | 0.3615 | 0.328 | 0.2858 | 0.2103 | 0.279 |
| **Year** | 2006 | 2011 | 2013 | 2014 | 2014 | 2014 | 2015 | 2013 | 2012 | 2013 | 2013 | 2009 |
| **Min** | 0.204 | 0.241 | 0.272 | 0.291 | 0.281 | 0.36 | 0.3475 | 0.3315 | 0.25 | 0.20893 | 0.167 | 0.182 |

Table 3: Monthly average of time series absolute difference between MODIS and MISR

| Year/Month | Jan. | Feb | Mar. | April | May | June | July | Aug. | Sep. | Oct. | Nov. | Dec. |
|---|---|---|---|---|---|---|---|---|---|---|---|---|
| 2006 | 0.026 | 0.005 | 0.01393 | 0.017 | 0.028 | 0.019 | 0.0023 | 0.0277 | 0.0307 | 0.008 | 0.001 | 0.0087 |
| 2007 | 0.031 | 0.019 | 0.024 | 0.017 | 0.012 | 0.016 | 0.0172 | 0.034 | 0.0202 | 0.003 | 0.002 | 0.0038 |
| 2008 | 0.002 | 0.048 | 0.00398 | 0.008 | 0.047 | 0.014 | 0.0099 | 0.0211 | 0.0239 | 0.039 | 0.005 | 0.0052 |
| 2009 | 0.007 | 0.028 | 0.02488 | 0.024 | 0.022 | 0.008 | 0.0158 | 0.0295 | 0.0276 | 0.01 | 0.004 | 0.0114 |
| 2010 | 0.01 | 0.002 | 0.04255 | 0.029 | 0.024 | 0.02 | 0.0025 | 0.0314 | 0.0376 | 0.014 | 0.019 | 0.0042 |
| 2011 | 0.01 | 0.003 | 0.03624 | 0.015 | 0.038 | 0.045 | 0.0126 | 0.0339 | 0.032 | 0.005 | 0.006 | 0.0076 |
| 2012 | 0.002 | 0.018 | 0.05673 | 0.015 | 0.033 | 0.006 | 0.0181 | 0.0296 | 0.0361 | 0.003 | 0.01 | 0.0044 |
| 2013 | 0.001 | 0 | 0.01323 | 0.021 | 0.032 | 0.007 | 0.0027 | 0.0207 | 0.0342 | 0.006 | 0.011 | 0.0022 |
| 2014 | 0.004 | 0.02 | 0.00132 | 0.015 | 0.037 | 0.019 | 0.0328 | 0.0412 | 0.0462 | 0.012 | 0.002 | 0.0048 |
| 2015 | 0.046 | 0.016 | 0.00905 | 0.029 | 0.038 | 0.014 | 0.0252 | 0.0406 | 0.0454 | 0.018 | 0.007 | 0.0406 |

SeaWiFS shows six years (2005 – 2010) of AOD spatial distribution with 0.5 degree (see Figure 3). The spatial map shows the same agreement in terms of the dust sources that shown by both MISR and MODIS.

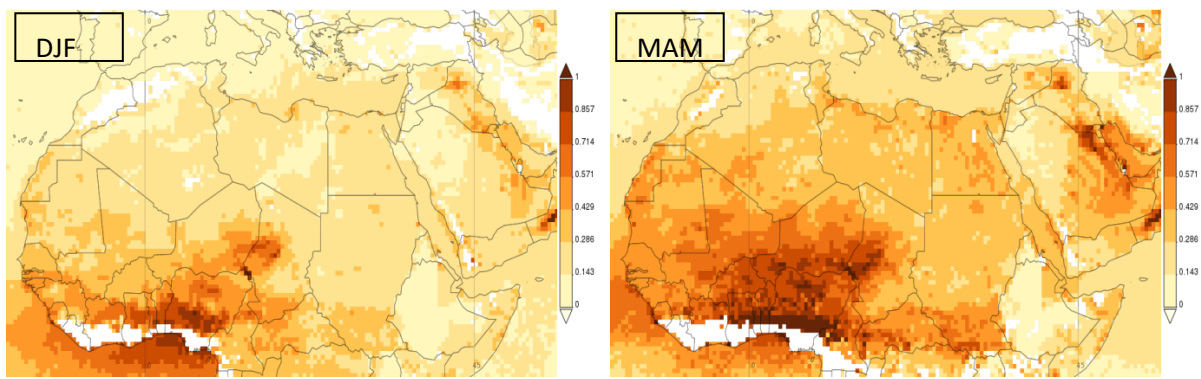

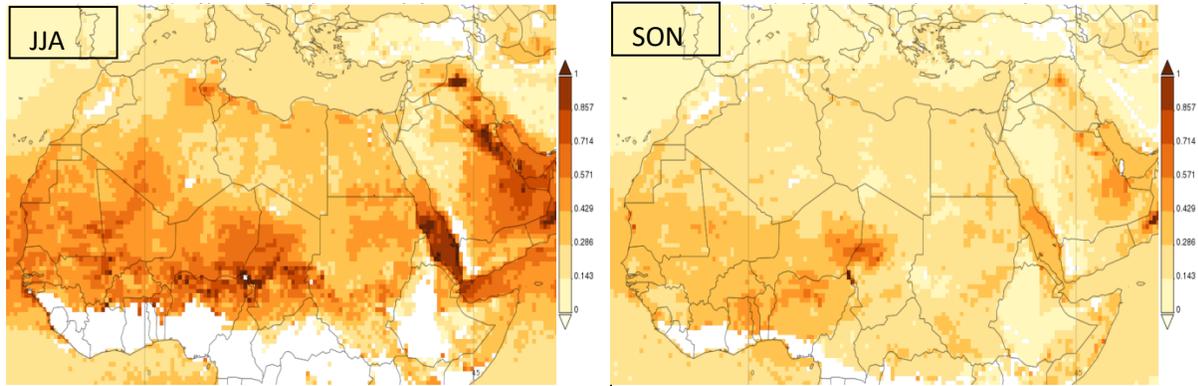

Figure 3: Six years (2005 – 2010) seasonal average of spatial distribution of AOD over MENA domain

Daily average time series between MODIS, MISR, SeaWiFS and OMI over MENA domain is shown in Figure 4. In general, the first three sensors estimate nearly the same AOD temporal pattern, the four sensors provide almost near daily average. On the other hand, OMI overestimates the AOD frequently, particularly during the severe events. The estimated average AOD at 19 March 2010 from MODIS, MISR, and OMI is 0.717, 0.5225, and 1.788 respectively. OMI overestimated this anomaly by more than twice MODIS and three times of MISR. The statistical quantities for the time series are identified in Table 4. The data in table 4 confirms that four sensors provide nearly the same data with slight differences due to errors which are more notable in the data provided through OMI, which resulted in higher average, variance and std deviation.

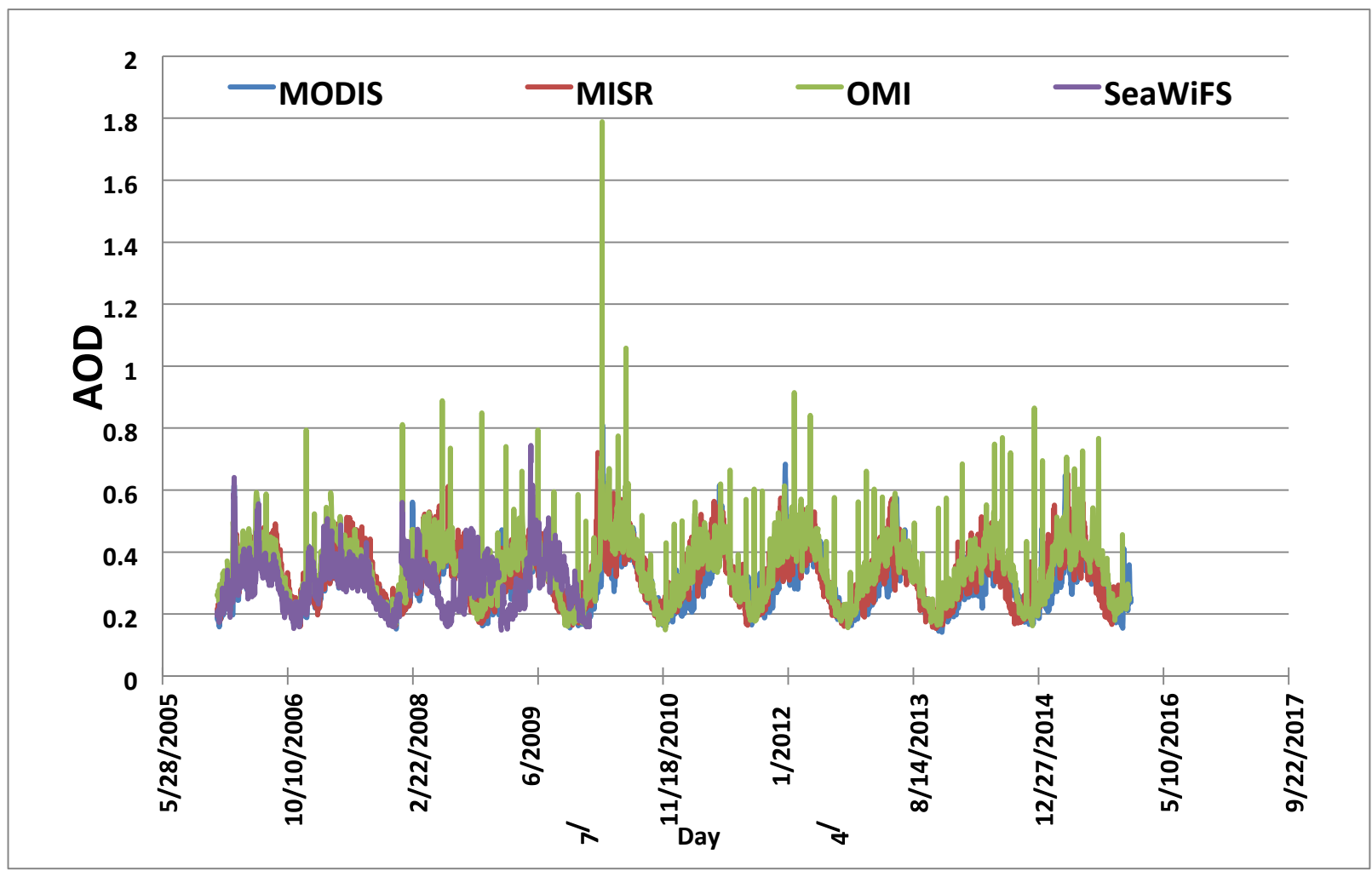

Figure 4: Ten years (2005-2016), Daily averages of time series from MODIS, MISR, OMI and SeaWiFS over MENA domain

The max and mini. observed AOD monitored by MODIS in the ten years were at 21 March 2010 and 12 June 2013 (0.8115 and 0.1414 ± 0.0862) respectively while from MISR were at 23 March 2010 and 12 April 2013 (0.721 and 0.153 ± 0.0906) respectively. The max AOD from OMI and SEAWIFS were 1.788 ± 0.1057 and 0.743 ± 0.082 respectively. In Conclusion the four sensors have the capability to capture the severe AOD event that occurred from 20 to 24 March 2010.

Table 4: AOD Statistics for presented Time series on Figure 4. Numbers in parentheses represent the corresponding date.

|  | Max. | Min. | Average | Variance | St. Deviation |
|---|---|---|---|---|---|
| **MODIS** | 0.81115 (3/21/2010) | 0.14148 (12/6/2013) | 0.3017 | 0.00743 | 0.0862 |
| **MISR** | 0.721335 (3/23/2010) | 0.15333 (12/4/2013) | 0.3256 | 0.00821 | 0.09064 |
| **OMI** | 1.78839 (3/24/2010) | 0.149049 (12/2/2010) | 0.3443 | 0.01118 | 0.10575 |
| **SEAWIFS** | 0.743847 (3/20/2010) | 0.148379 (11/11/2009) | 0.2993 | 0.00672 | 0.08203 |

### 3.2 Sulfate Pollutants ($SO_2$ and $SO_4$)

The daily average mass concentration trend of $SO_2$ and $SO_4$ for 10 years (2005-2015) over the MENA domain are shown in figure (5). it is shown that the max. Mass concentration for $SO_2$ takes place in December 2015 with a value of 1.48138E-09 ± 1.086E-10 and the max. Mass concentration for $SO_4$ at August 2008 with value 1.81E-09 ± 1.827E-10. However the Min. mass concentration for both $SO_2$ and $SO_4$ occur at (July 2014 and March 2013) with values 9.45256E-10 ± 1.827E-10 and 9.92E-10 ± 1.82788E-10 respectively. There is no significant trend for both pollutants shown from this time series trend. Table (5) summarizes the statistical indicators for both pollutants time series.

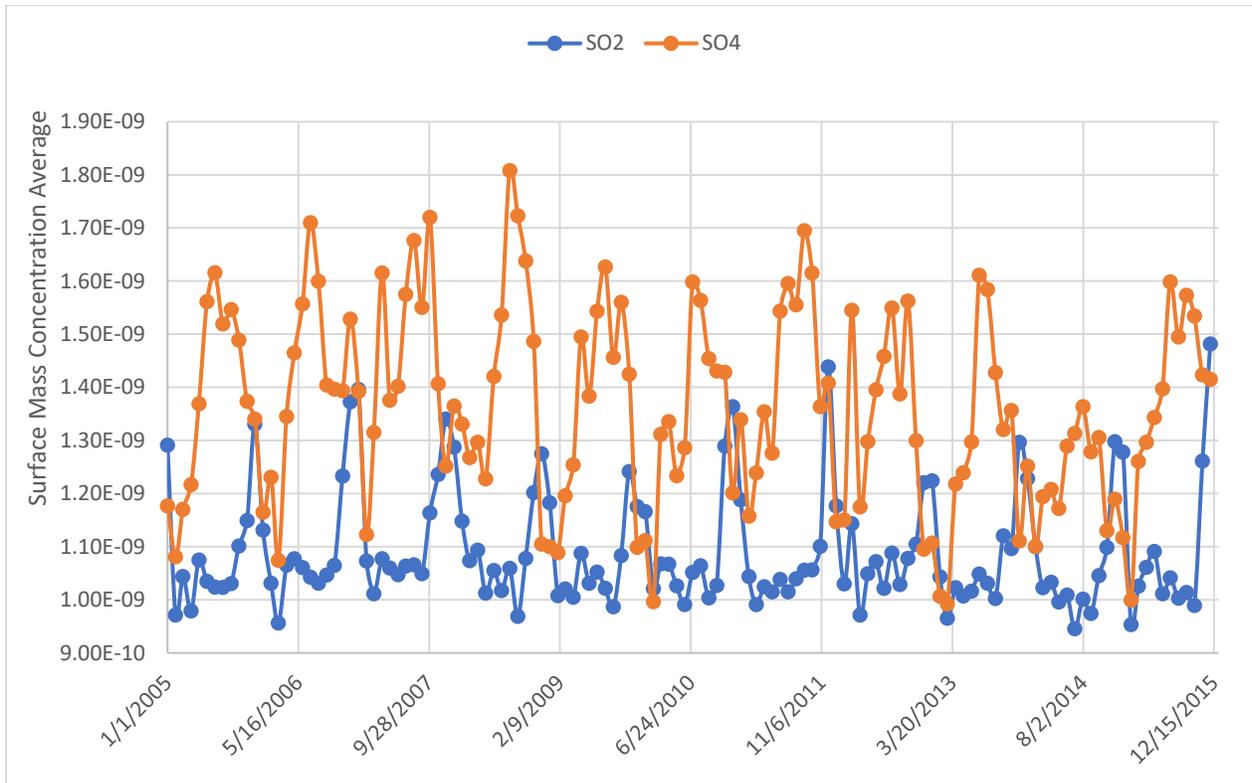

Figure (5): Daily time series of So2 and So4 over the MENA Domain

Table (5): Statistics for the presented daily time series of So2 and So4 over the MENA domain in Figure (5)

|  | SO2 | SO4 |
| --- | --- | --- |
| Max | 1.48138E-09 (Dec-15) | 1.81E-09 (Aug-08) |
| Min | 9.45256E-10 (Jul-14) | 9.92E-10 (Mar-13) |
| Average | 1.08998E-09 | 1.36241E-09 |
| Std. Deviation | 1.08682E-10 | 1.82788E-10 |
| Variance | 1.18118E-20 | 3.34115E-20 |

Interannual surface mass concentration for both $SO_2$ and $SO_4$ over the MENA region is shown in figure (6) and Figure (7). The data obtained from MERRA-2 shows that the highest concentration of SO2 notably occurs in winter especially in December with a maximum of 1.48E-09 in December 2015 and the highest average across the 10 years span of 1.327E-09. Whilst in figure (7) and Table (7) that the highest average concentration of $SO_4$ takes place in August with the highest peak at 2008 with a value of 1.81E-09 and with the highest average with a value of 1.58E-09.

From the slopes in the figure (6) and table (6). it's notable that there is a high tendency of increase in SO2 mass concentration in the last 2 years throughout the winter months, whilst in figure (7) it's notable that most of the values fluctuate within the same range, although it shows an increase in the months (July … October) over the last 2 years as well which is aligned with obtained results before that revealed the highest concentrations of particulate sulfate and nitrate and gaseous nitric acid were found in the summer season over urban areas(Dokki , Giza, Egypt) when measured in the winter season (1999–2000) and summer season (2000) (Khoder,2002)

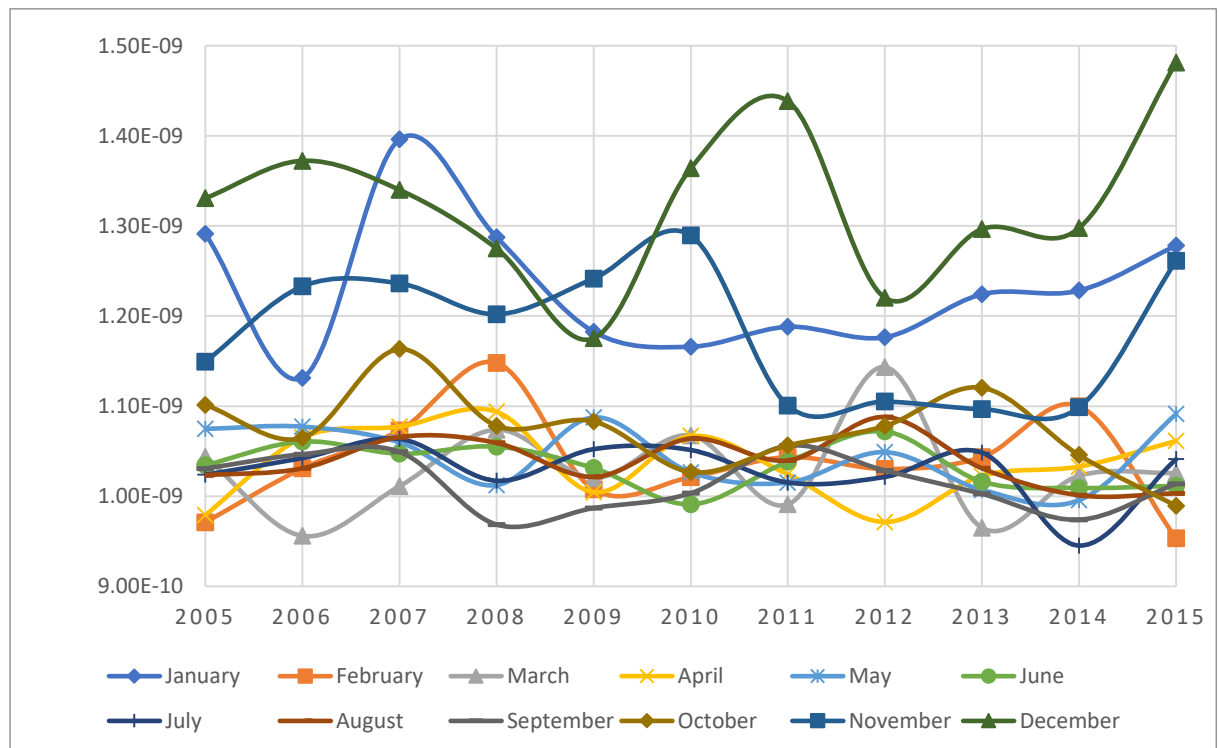

Figure 6: Interannual monthly average surface mass concentration for both $SO_2$ over the MENA region

Table 6: So$_2$ Statistics for presented time series on Figure 6

| Month | Max | Min | Average | Variance | Std. Deviation |
|---|---|---|---|---|---|
| January | 1.4E-09 | 1.13E-09 | 1.232E-09 | 5.7639E-21 | 7.59206E-11 |
| February | 1.15E-09 | 9.53E-10 | 1.038E-09 | 3.0199E-21 | 5.4954E-11 |
| March | 1.14E-09 | 9.56E-10 | 1.029E-09 | 2.8126E-21 | 5.30343E-11 |
| April | 1.09E-09 | 9.71E-10 | 1.036E-09 | 1.62E-21 | 4.02497E-11 |
| May | 1.09E-09 | 9.96E-10 | 1.045E-09 | 1.2295E-21 | 3.50637E-11 |
| June | 1.07E-09 | 9.91E-10 | 1.033E-09 | 6.061E-22 | 2.46191E-11 |
| July | 1.06E-09 | 9.45E-10 | 1.029E-09 | 1.0362E-21 | 3.21905E-11 |
| August | 1.09E-09 | 1E-09 | 1.039E-09 | 7.5522E-22 | 2.74812E-11 |
| September | 1.06E-09 | 9.69E-10 | 1.015E-09 | 9.1952E-22 | 3.03237E-11 |
| October | 1.16E-09 | 9.89E-10 | 1.073E-09 | 2.1581E-21 | 4.64551E-11 |
| November | 1.29E-09 | 1.1E-09 | 1.183E-09 | 5.5081E-21 | 7.42168E-11 |
| December | 1.48E-09 | 1.18E-09 | 1.327E-09 | 7.8751E-21 | 8.8742E-11 |

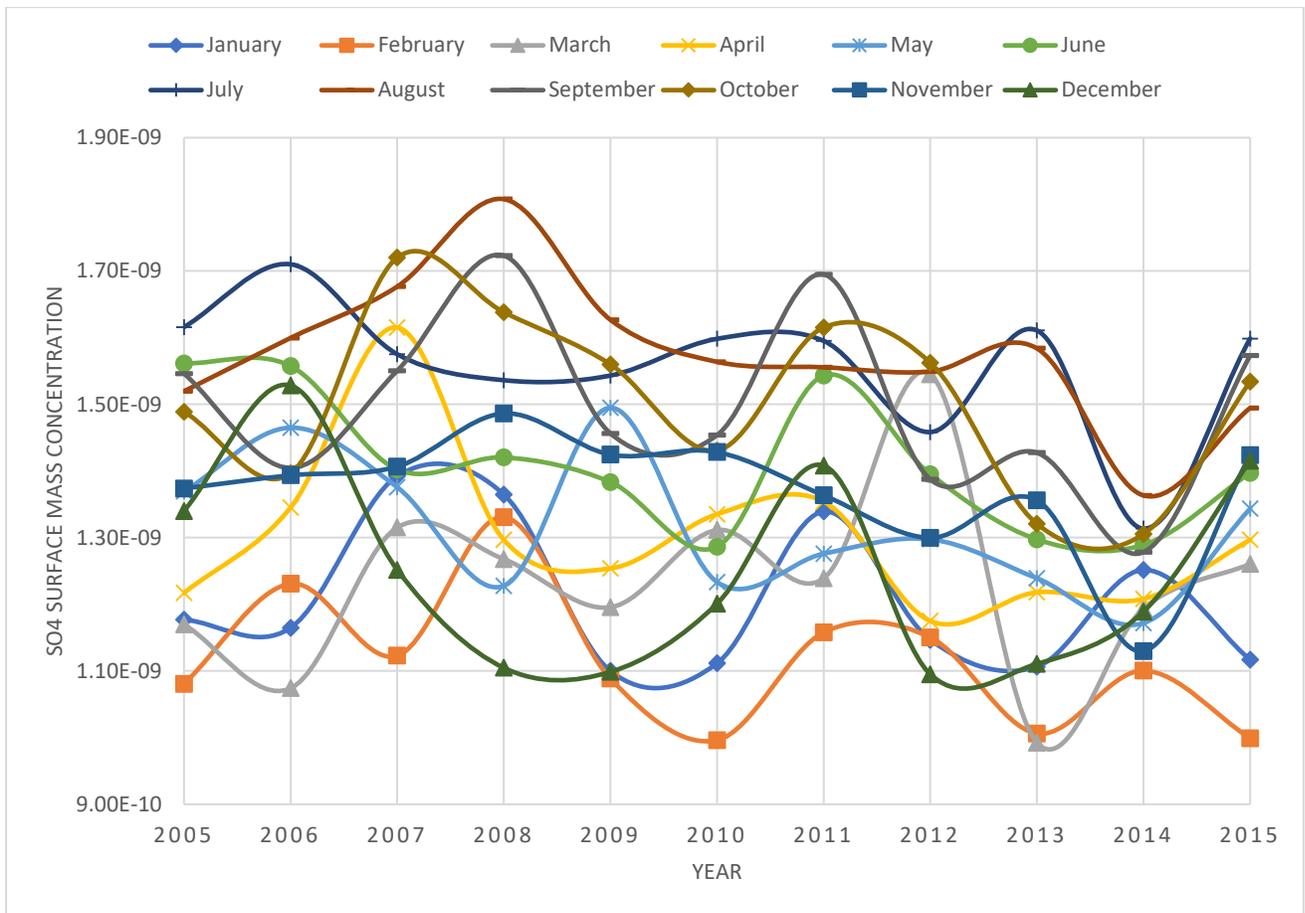

Figure 7: Interannual monthly average surface mass concentration for both $SO_4$ over the MENA region

Table 7: $So_4$ Statistics for presented time series on Figure 6

| Month | Max | Min | Average | Variance | Std. Deviation |
|---|---|---|---|---|---|
| January | 1.39E-09 | 1.10E-09 | 1.21E-09 | 1.24E-20 | 1.11E-10 |
| February | 1.33E-09 | 9.96E-10 | 1.11E-09 | 1.05E-20 | 1.02E-10 |
| March | 1.55E-09 | 9.92E-10 | 1.23E-09 | 2.02E-20 | 1.42E-10 |
| April | 1.62E-09 | 1.18E-09 | 1.30E-09 | 1.45E-20 | 1.21E-10 |
| May | 1.49E-09 | 1.17E-09 | 1.32E-09 | 1.05E-20 | 1.02E-10 |
| June | 1.56E-09 | 1.29E-09 | 1.41E-09 | 1.06E-20 | 1.03E-10 |
| July | 1.71E-09 | 1.31E-09 | 1.56E-09 | 1.05E-20 | 1.02E-10 |
| August | 1.81E-09 | 1.36E-09 | 1.58E-09 | 1.23E-20 | 1.11E-10 |
| September | 1.72E-09 | 1.28E-09 | 1.50E-09 | 1.78E-20 | 1.33E-10 |
| October | 1.72E-09 | 1.31E-09 | 1.51E-09 | 1.75E-20 | 1.32E-10 |
| November | 1.49E-09 | 1.13E-09 | 1.37E-09 | 8.75E-21 | 9.35E-11 |
| December | 1.53E-09 | 1.10E-09 | 1.25E-09 | 2.31E-20 | 1.52E-10 |

The spatial maps of $SO_2$ surface mass concentration of the 4 seasons throughout the ten years is shown in the Figure (8). The spatial maps from reanalysis MERRA-2 data of the four season's shows that the areas with the highest concentrations are the delta of Egypt, Kuwait, the eastern south cities of Iraq, which is aligned with the locations of highest emissions of So2 over the spatial global maps that were built based on OMI satellite. These maps showed decrease on the So2 emissions over Delta, Egypt by around 25 % between 2005-2007 and 2012-2014. Over eastern south Iraq, the emissions increased around 25 % however, over Kuwait there is general decrease by different percentage 25 %, 50 % based on the location inside Kuwait (Vitali E. Fioletov,..etal 2016).

On the other hand, the spatial concentration of So4 is shown on the seasonal spatial in Figure (9). Generally, the concentration of So4 over the Mediterranean Sea is high at both spring and summer seasons while over Red Sea, it is shown relatively high concentration in all seasons especially in the summer season. The area in Box A (shown in upper right of Figure: 9) is also identified as high concentration region at all seasons. This domain includes Arabian Gulf, eastern of Saudi Arabia, UAE, Qatar and eastern south of Iraq. Over Egypt highest concentration areas is different based on the season, with continuous high concentration of So4 over Delta in the four seasons. It is shown high concentration of So4 over north of Egypt especially over the Delta and great Cairo over the summer although in autumn that half of eastern Egypt shows high concentration of So4 also.

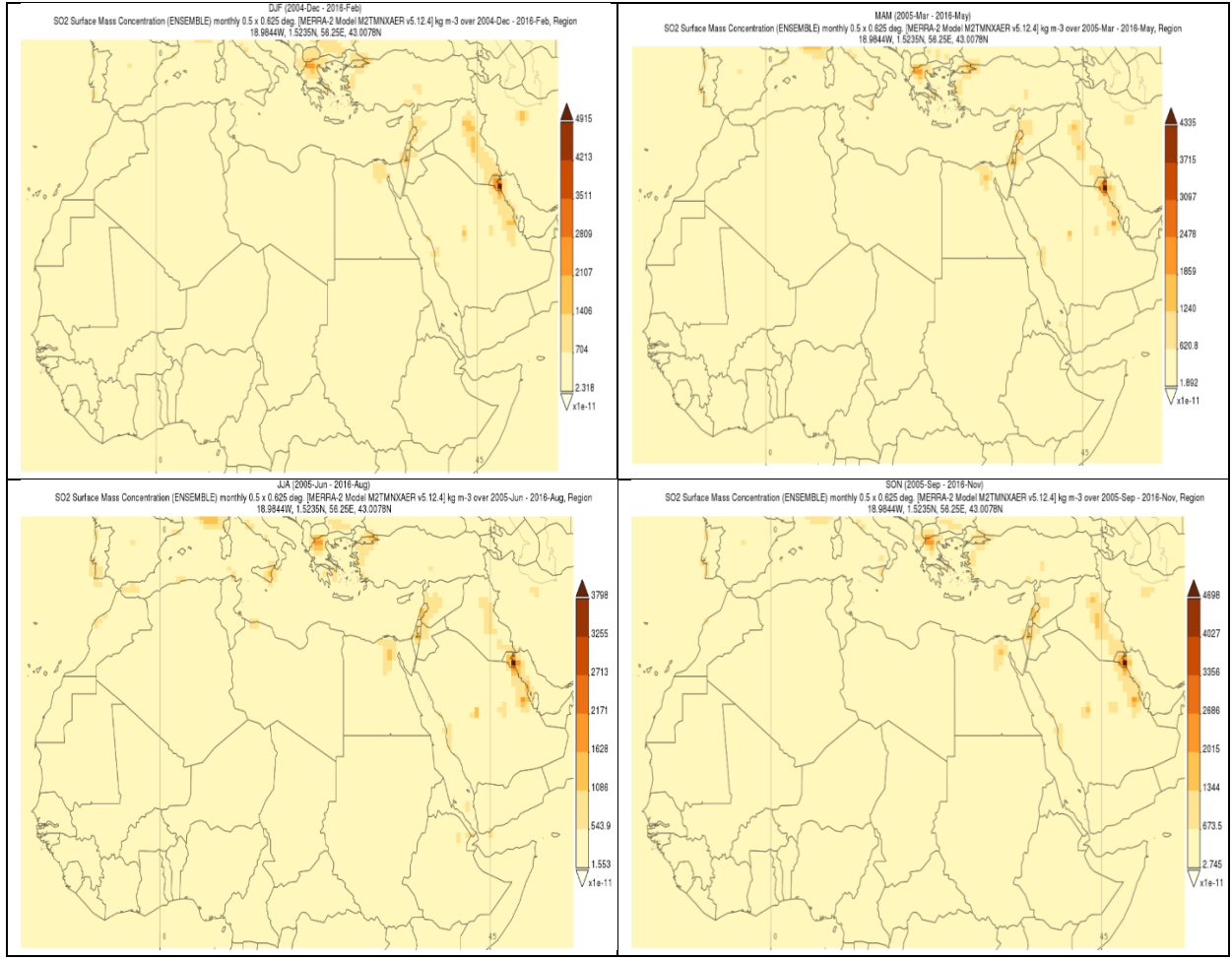

Figure (8): Seasonal Spatial distribution of So2 over the MENA Domain

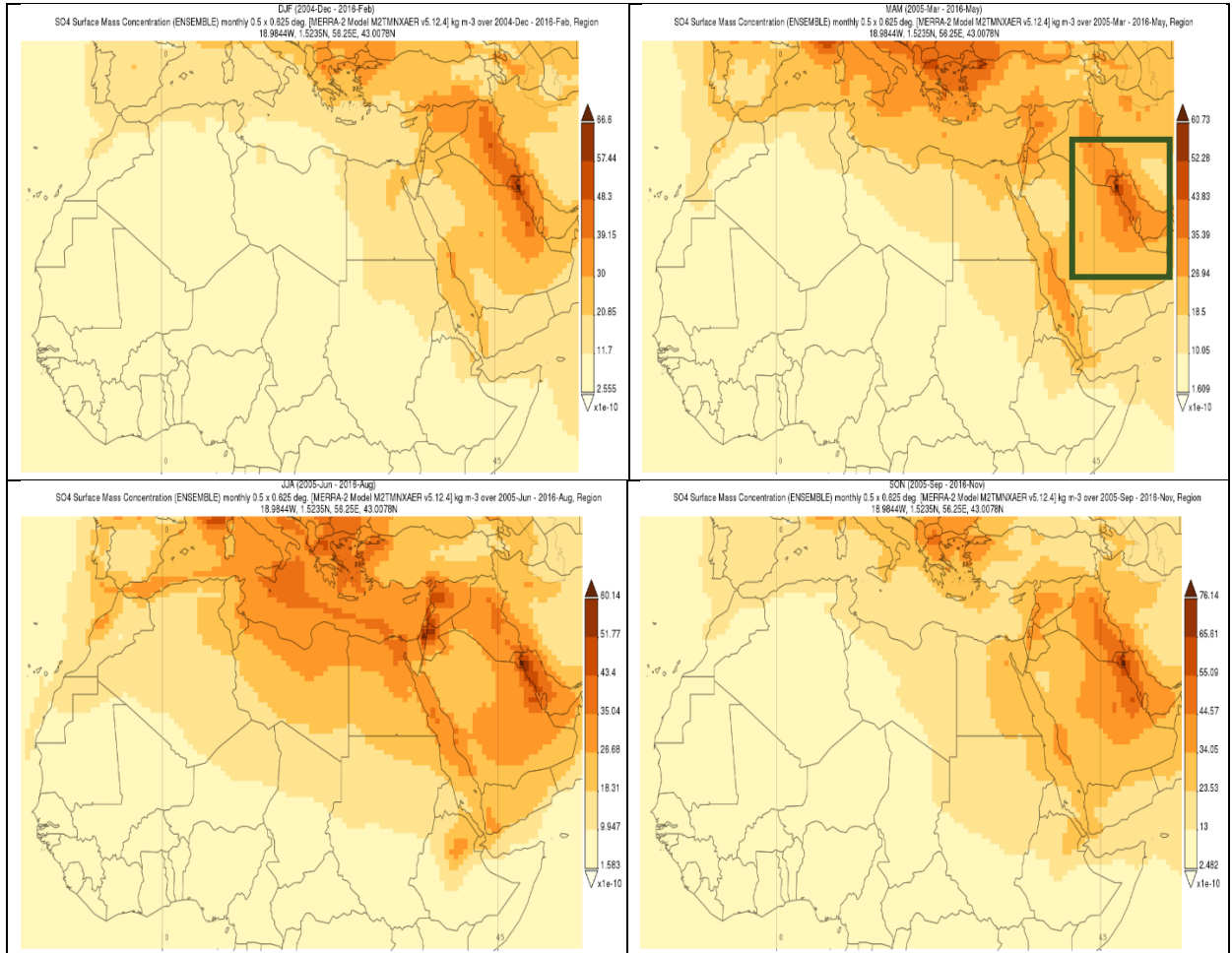

Figure (9): Seasonal Spatial distribution of $So_4$ over the MENA Domain, Region A identified as the black box on the right upper box

Figure (10) shows the statistical indicators for the seasonal spatial distribution of $So_2$ [Upper Figure] and $So_4$ [Lower Figure] over the MENA domain. It is shown that the winter has highest average surface mass concentration with value 1.2E-9. On the other hand, the highest average surface mass concentration of $So_4$ in the summer season with average 1.55 E-9. This is result in high temperature summer season which leads to high emissions of water particle. The minimum average mass concentration of $So_4$ and $So_2$ occur in the winter and in spring with values 1.24 E-9 and 1.05E-9.

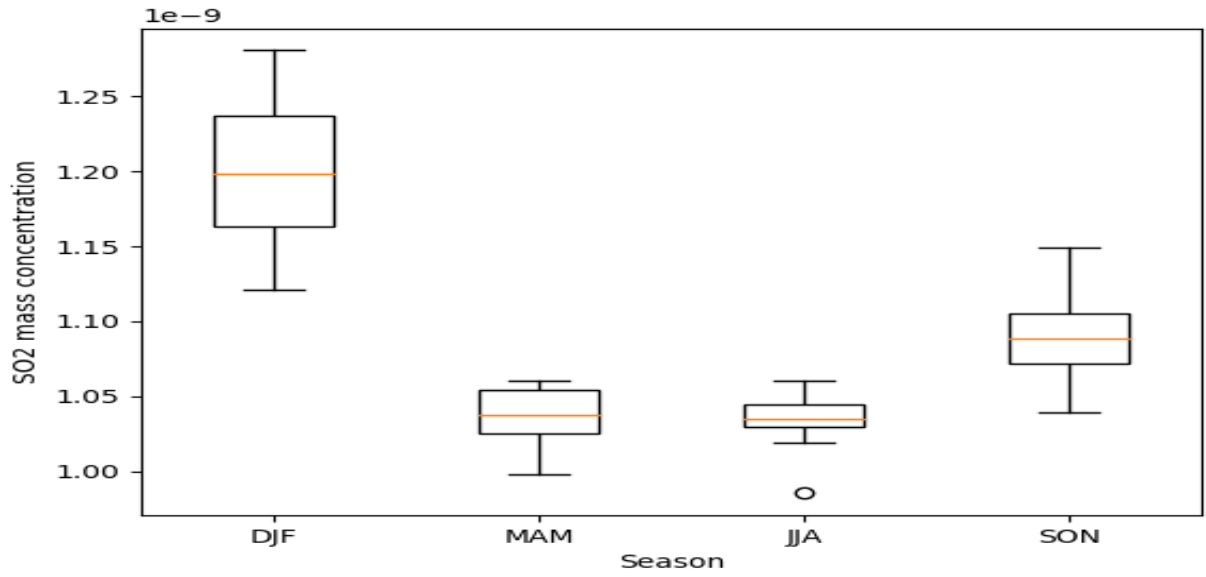

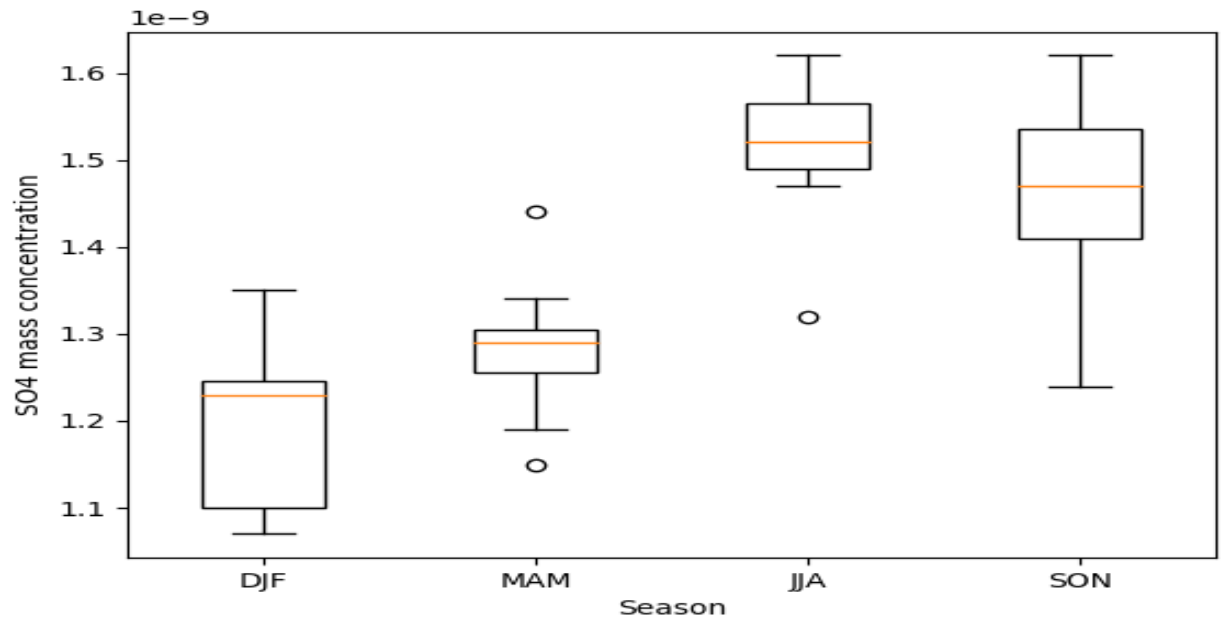

Figure (10): Box plot for the statistical values for the seasonal $So_2$ [Upper panel] and $So_4$ [Lower Panel].

## 4- Conclusions and Future Work

In this Current Study, long temporal trend of AOD for ten years (2005-2016) from four satellite sensors MODIS, MISR, OMI and SEAWIFS were established. For the ten years MODIS and MISR almost generated the same seasonal spatial maps. The two sensors have the capability to capture the highest four active dust sources. In terms of Interannual seasonal time series, MISR tends to underestimate AOD in comparison to MODIS over the MENA domain throughout the four seasons. Although SEAWIFS data were available only for six years over the MENA domain, they were sufficient to show its capabilities in capturing the same dust sources. The Four sensors have the capability to capture highly severe AOD event over the MENA Domain.

The Spatiotemporal analysis for two air Pollutants So2 and So4 over MENA domain was conducted using MERRA-2 reanalysis data. $SO_4$ has very high concentration over the Gulf Arabia region and its surrounding area in all four seasons. In addition to always-high concentration over water areas such the Mediterranean and Red Seas in all seasons, especially on the summer season. SO2 Shows high concentration over Nile Delta (Egypt) , Kuwait and parts of Iraq over the four seasons.

This current research can be considered as guide on the future research which include, working on Numerical modelling, chemical transport model (CTM) to predict the physical indicator AOD and different air pollutants (So2, So4, Co, Co4,…etct) and their generative chemical mechanisms, their concentration over MENA domain.

38-
Fioletov, V. E., McLinden, C. A., Krotkov, N., Li, C., Joiner, J., Theys, N., Carn, S., and Moran, M. D.: A global catalogue of large $SO_2$ sources and emissions derived from the Ozone Monitoring Instrument, Atmos. Chem. Phys., 16, 11497-11519, https://doi.org/10.5194/acp-16-11497-2016, 2016.